%% file: main.tex
\documentclass[runningheads]{llncs}
\usepackage[T1]{fontenc}
\usepackage{framed} 
\usepackage{hyperref}
\usepackage{longtable}
\usepackage{multirow}
\usepackage{graphicx}
\usepackage{subcaption}
\usepackage[table,xcdraw]{xcolor}
\usepackage[normalem]{ulem}
\usepackage{amsfonts} 
\usepackage{amsmath}
\usepackage{booktabs} 
\usepackage{enumitem}
\usepackage[bottom]{footmisc}

\makeatletter
\def\blfootnote{\xdef\@thefnmark{}\@footnotetext}
\makeatother

\begin{document}
%
\title{Pre-trained LLMs Meet Sequential Recommenders: Efficient User-Centric Knowledge Distillation}
\titlerunning{Pre-trained LLMs Meet Sequential Recommenders...}

\author{
Nikita Severin\inst{1} \and Danil Kartushov\inst{5} \and Vladislav Urzhumov\inst{3} \and Vladislav Kulikov\inst{5} \and Oksana Konovalova\inst{3} \and Alexey Grishanov\inst{2} \and Anton Klenitskiy\inst{2} \and Artem Fatkulin\inst{2} \and Alexey Vasilev\inst{2,4} \and Andrey Savchenko\inst{2,4}
\and Ilya Makarov\inst{6}
}

\authorrunning{N. Severin et al.}

\institute{
Independent researcher, Belgrade, Serbia\and
Sber AI Lab, Moscow, Russia\and
Innopolis University, Innopolis, Russia\and
HSE University, Moscow, Russia\and
ITMO University, Saint Petersburg, Russia\and
AIRI, Moscow, Russia
\\
\email{nseverin14@gmail.com, grishanov.av@phystech.edu}
}

%
%
%
\maketitle


\input{sections/0_abstract}
\input{sections/1_intro}
\input{sections/2_approach}

\input{sections/3_experimental_setup}
\input{sections/4_results}

\input{sections/5_conclusion}

\begin{credits}
\subsubsection{Disclosure of Interests.} 
The authors have no competing interests to declare that are relevant to the content of this article.
\end{credits}

\subsubsection*{Publisher's Note}
{\scriptsize This version of the contribution has been accepted for publication, after peer review but is not the Version of Record and does not reflect post-acceptance improvements, or any corrections. The Version of Record is available online at: \url{http://dx.doi.org/10.1007/978-3-032-21300-6_42}. Use of this Accepted Version is subject to the publisher's Accepted Manuscript terms of use: \url{https://www.springernature.com/gp/open-research/policies/accepted-manuscript-terms}.}

\bibliographystyle{splncs04}
\bibliography{sample-base}

\end{document}

%% file: sections/0_abstract.tex
\begin{abstract}
Sequential recommender systems have achieved significant success in modeling temporal user behavior but remain limited in capturing rich user semantics beyond interaction patterns. Large Language Models (LLMs) present opportunities to enhance user understanding with their reasoning capabilities, yet existing integration approaches create prohibitive inference costs in real time. To address these limitations, we present a novel knowledge distillation method that utilizes textual user profile generated by pre-trained LLMs into sequential recommenders without requiring LLM inference at serving time. The resulting approach maintains the inference efficiency of traditional sequential models while requiring neither architectural modifications nor LLM fine-tuning. 

\keywords{Recommender system \and Large Language Model (LLM) \and sequential recommendation \and knowledge distillation.}

\end{abstract}

%% file: sections/1_intro.tex
\section{Introduction}

Sequential recommender systems (SRS) have emerged as essential components for modeling temporal user behavior and delivering personalized recommendations. While transformer-based architectures like SASRec~\cite{kang2018self} and BERT4Rec~\cite{sun2019bert4rec} have advanced the field significantly, these systems continue to face fundamental challenges: data sparsity leading to poor generalization and limited ability to capture
user semantics beyond interaction patterns~\cite{koren2021advances,kumar2019predicting}.

The rise of Large Language Models (LLMs) offers promising opportunities to enhance recommendation systems through their sophisticated semantic understanding capabilities~\cite{bao2023bi,lin2023sparks,wei2022emergent,zhao2023survey}. This has led to methods ranging from zero-shot prompting~\cite{bao2023bi,hou2023llamarec,hou2024llmrank,lin2023sparks,wei2022emergent,zhao2023survey} and feature augmentation~\cite{chen2023tbin,liu2022boosting,liu2021pre,sun2024llmcf,xi2023towards,zhang2022twhin} to full LLM fine-tuning for recommendation tasks~\cite{geng2022recommendation,raffel2020exploring,tan2024idgenrec}. However, these approaches face critical limitations for practical deployment including high-latency LLM inference during serving. Recent efforts to distill LLM knowledge into recommender models~\cite{severin2024llm} often remain item-centric and  require expensive LLM fine-tuning ~\cite{li2023e4sreceleganteffectiveefficient,10.1145/3640457.3688121,Liu_Wu_Wang_Wang_Zhu_Zhao_Tian_Zheng_2025,sun2024llmcf,tian2024reland}, failing to fully leverage user-specific semantics without incurring high inference costs.

We introduce a novel knowledge distillation method that transfers user-centric knowledge from pre-trained LLMs into sequential recommenders without altering their architecture or requiring LLM inference at serving time. It builds on the intuition that effective recommendation models must implicitly reconstruct rich user-specific information within their internal representations. During training, the model aligns these representations with LLM-generated textual user profiles, embedding semantic information directly into its parameters. Our two-phase training first distills user profile knowledge through an auxiliary reconstruction loss, then fine-tunes exclusively on the recommendation objective.

To the best of our knowledge, this is the first work to distill knowledge from a pre-trained LLM without domain-specific fine-tuning into sequential recommendation models.
Extensive experiments across four benchmark datasets confirm that our method significantly enhances recommendation quality while preserving computational efficiency.

%% file: sections/2_approach.tex
\section{Proposed Approach}
Let $\mathcal{U} = \{ u_1, u_2, \dots, u_N \}$ be the set of users, and $\mathcal{I} = \{ i_1, i_2, \dots, i_M \}$ be the set of items. In the context of sequential recommendation systems, for each user $u \in \mathcal{U}$ we are given an ordered interaction history 
$\mathcal{S}_u = (s_t^u \mid t = 1, \dots, |\mathcal{S}_u|)$. 
The task is to predict the next item a user will interact with.
\begin{figure}[t]
  \centering
  \includegraphics[width=\linewidth]
  {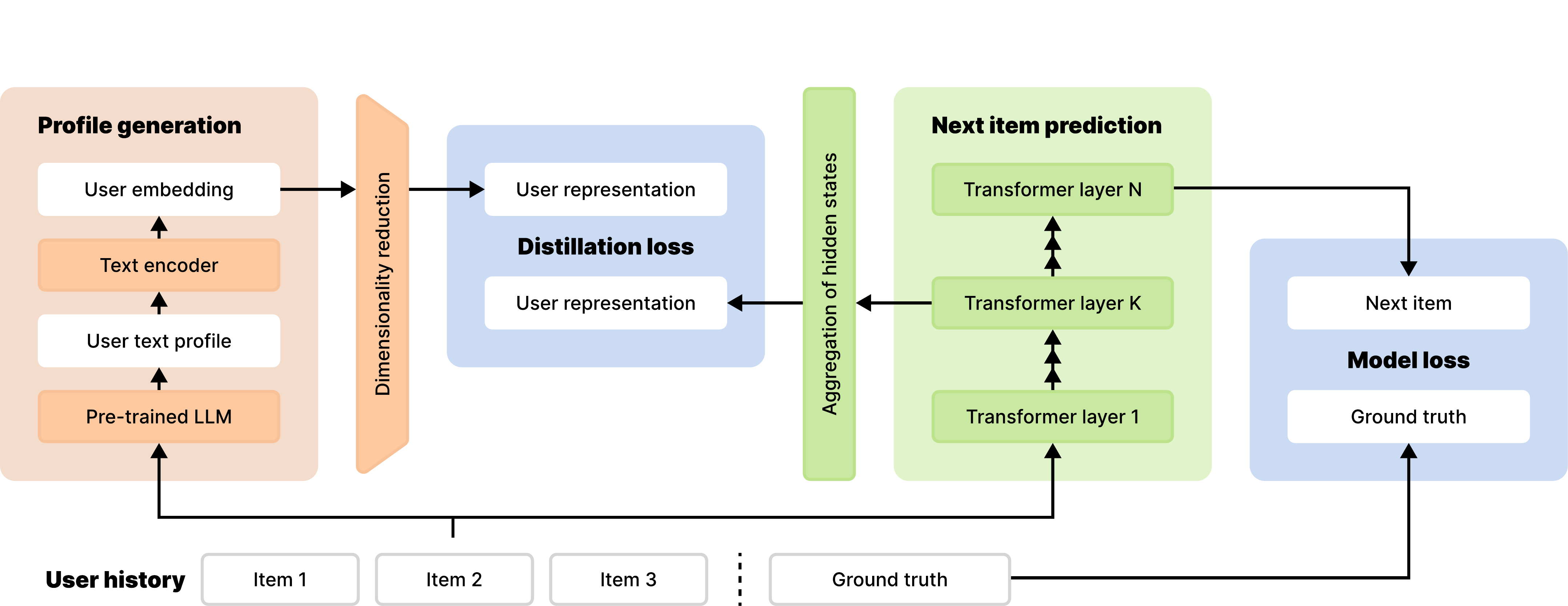}
\caption{Proposed knowledge transfer approach from LLM to a Transformer-based sequential recommendation model.}
  \label{fig1:pipeline}
\end{figure}

In this paper we propose a lightweight method (Figure~\ref{fig1:pipeline}) to distill LLM knowledge for this problem. Our method consists of the following steps.

\textbf{Obtaining LLM-based user representation.} Given user $u$, we first aggregate available textual metadata from their interaction history $\mathcal{S}_u$ to describe the user's behavioral patterns in plain text. The specific metadata varies by dataset: examples include item titles, categories, descriptions for products; titles, genres, and plot summaries for movies; and other domain-relevant attributes.

This aggregated text is then fed to a pre-trained LLM with a carefully designed prompt to generate a comprehensive textual user profile $P(u)$ (Figure~\ref{fig2:llm_profiles}). Our prompt design follows a consistent structure across datasets, instructing the LLM to: (1)-(2) analyze the user's interaction history, (3) identify key preferences and patterns, (4) distinguish highly-rated from poorly-rated items, and (5) synthesize an overall user characterization. While the prompt structure remains consistent, specific details are optimized for each dataset's domain.

Next, we obtain vector representation $T(E(P(u)))$ based on the generated profile, employing the textual encoder
$E$ followed by transformation $T$ projecting these embeddings into the same space as recommender models operate with. We employed UMAP \cite{mcinnes2018umap} as $T$ to reduce dimensions while preserving user-to-user intrinsic relationships and distances. These profile embeddings $T(E(P(u)))$ are pre-computed once and remain frozen throughout training, serving as fixed targets for the distillation objective.

\begin{figure}[t]
\centering
\fbox{
\begin{minipage}{1\linewidth}
\small
\begin{enumerate}[leftmargin=*, itemsep=2pt, topsep=2pt]
    \item This user displays a clear preference for skincare and makeup products, with a notable interest in nail care as well~\dots
    \item The user expresses a clear appreciation for natural and organic ingredients, as evidenced by~\dots
    \item Highly-rated products share commonalities in terms of~\dots
    \item Products receiving low ratings appear to lack~\dots
    \item Overall, this user appears to be a discerning beauty enthusiast who prioritizes natural, high-quality products that deliver visible results~\dots\ They would appreciate recommendations for organic skincare brands and innovative nail polish shades~\dots
\end{enumerate}
\end{minipage}
}
  \caption{Example of Beauty user profile inferred from LLM.}
  \label{fig2:llm_profiles}
\end{figure}

\textbf{Distillation of LLM-based user profiles into sequential recommender} is achieved through a two-stage training strategy:

\textit{1. Distillation stage.} During the first stage, the model is optimized using both the next item prediction and the auxiliary distillation losses to align sequential recommender with LLM-based user representation $T(E(P(u)))$.

Let $h_t^k \in \mathbb{R}^d$ denote the representation of interaction $s_t$ in the interaction history $\mathcal{S}_u$ obtained after the $k$-th block of the transformer-based recommender model. Aggregating these representations across the context length, we define $H_k(\mathcal{S}_u) \in \mathbb{R}^d$ as the embedding of the entire interaction sequence $\mathcal{S}_u$.

To obtain $H_k(\mathcal{S}_u)$, we extract hidden states from $k$-th layer of the transformer and apply an aggregation strategy. Two pooling strategies are employed depending on dataset characteristics: mean pooling $H_k = \frac{1}{m}\sum_{t=1}^{m} h_t^k$ for datasets with stable user preferences or exponential weighting $H_k = \sum_{t=1}^{m} \left( \frac{\exp(\gamma \cdot t)}{\sum_{j=1}^{m} \exp(\gamma \cdot j)} \right) \cdot h_t^k$ for datasets where recent interactions are more predictive. In the exponential weighting case, the hyperparameter $\gamma$ controls the emphasis on recency; higher values place more weight on recent interactions.

Auxilary distillation loss $\mathcal{L}_{distill}(u, k) = \mathcal{L}_{mse}(H_k(\mathcal{S}_u),\ T(E(P(u))))$ is computed as MSE between LLM-based and sequential recommender-based user representations.
\( \mathcal{L}_{model} \) represents the standard objective (for example, cross-entropy for next-item prediction). Total loss during the first stage:

\begin{equation} \label{eq:4}
    \mathcal{L} = \alpha \cdot \mathcal{L}_{distill} + (1 - \alpha) \cdot \mathcal{L}_{model}.
\end{equation}

Since \( \mathcal{L}_{distill} \) is often much smaller than \( \mathcal{L}_{model} \), a dynamic scaling factor $\beta$ is introduced. It is computed per batch to balance $\mathcal{L}_{distill}$ and $\mathcal{L}_{model}$ contributions:

\begin{equation} \label{eq:5}
    \mathcal{L} = \alpha \cdot \beta \cdot \mathcal{L}_{distill} + (1 - \alpha) \cdot \mathcal{L}_{model},
\end{equation}
\begin{equation} \label{eq:6}
    \beta = \operatorname{detach}\left(\frac{\mathcal{L}_{model}}{\mathcal{L}_{distill}}\right),
\end{equation}
where \( \operatorname{detach}(\cdot) \) 
operator denotes the stop-gradient operation, ensuring 
that \(\beta\) does not participate in gradient updates. The factor $\beta$ 
acts as a per-batch normalization coefficient that rescales the distillation loss relative to the model loss, preventing the smaller $\mathcal{L}_{distill}$ term from being 
numerically dominated. 

\textit{2. Fine-tuning stage.} During the second stage, the auxiliary task is removed, and the model is trained solely on the next-item prediction task with $\mathcal{L}_{model}$.

%% file: sections/3_experimental_setup.tex
\section{Experimental setup}

\begin{table}[t]
    \centering
    \caption{Statistics of the datasets}
    \label{tab0:commands}
    \resizebox{\textwidth}{!}{%
        \begin{tabular}{l c c c c c c c}
            \hline
            \textit{\textbf{Dataset}} & \textit{\textbf{Domain}} &
            \textit{\textbf{\#Users}} & \textit{\textbf{\#Items}} & \textit{\textbf{\#Interactions}} & \textit{\textbf{Avg. Length}} & \textit{\textbf{Density}} \\ \hline
            Beauty\cite{beauty} & Product Reviews  & 70,996 & 39,116 & 436,309 & 6.145 & 0.00304 \\ 
            ML20M\cite{movielens} & Movies & 137,165 & 13,132 & 19,933,088 & 143.929 & 0.01096 \\
            Kion\cite{kion} & Movies & 16,797 & 5,626 & 287,698 & 17.128 & 0.00055 \\ 
            Amazon M2~\cite{amazon_m2} & E-Commerce & 616,502 & 334,060 & 3,651,542 & 5.923 & 0.00002 \\
            \hline
        \end{tabular}
     }
\end{table}

\textbf{Datasets}. We have evaluated our method on four publicly available datasets (Table \ref{tab0:commands}) spanning different domains, scales, and levels of density. To ensure data quality, we applied k-core filtering with k=5 for all datasets, removing users and items with fewer interactions than this threshold. This cutoff was selected to balance data sparsity reduction with sufficient dataset coverage which is a standard practice \cite{10.1145/3640457.3688195,kion,tikhonovich2025esasrec}.

\textbf{Baselines}. To evaluate the effectiveness of the proposed approach, we compare it against two strong sequential recommendation baselines: SASRec~\cite{sasrec} with cross-entropy loss \cite{klenitskiy2023turning} and BERT4Rec~\cite{sun2019bert4rec}. Since our method distills knowledge from a pre-trained LLM, we include IDGenRec~\cite{tan2024idgenrec} as an LLM-based baseline. IDGenRec has demonstrated solid performance in recent work~\cite{tan2024idgenrec,wang2024towards,gram} 
by fine-tuning LLMs to generate semantic item identifiers, making it a strong comparison point for evaluating LLM-based recommendation approaches. 
It fine-tunes LLM and uses it during inference to generate semantic item identifiers and make predictions.

\textbf{Training details}. Following \cite{10.1145/3640457.3688195}, we adopt a global temporal split (80/20) to avoid test leakage \cite{10.1145/3569930,Time2Split}: interactions before the threshold form the training set, while later ones define the test set. 

User profiles for distillation are generated using the pre-trained Gemma-2-9b\footnote{\href{https://huggingface.co/UCLA-AGI/Gemma-2-9B-It-SPPO-Iter3}{https://huggingface.co/UCLA-AGI/Gemma-2-9B-It-SPPO-Iter3}} \cite{team2024gemma} and encoded using the multilingual E5-large model\footnote{\href{https://huggingface.co/intfloat/multilingual-e5-large}{https://huggingface.co/intfloat/multilingual-e5-large}} \cite{e5}, producing 1024-dimensional embeddings. These embeddings are reduced via UMAP to align with the sequential models' hidden state dimensionality.

For both SASRec and BERT4Rec, extensive hyperparameter tuning was performed using grid search over learning rates \{0.0001, 0.0005, 0.001\}, batch sizes \{128, 256, 512\}, dropout rates \{0.2, 0.3, 0.5\}, number of transformer layers \{1, 2, 4\} and number of heads in multi-head attention \{2, 4, 8\}.

For our approach, we additionally tuned over the loss weighting parameter $\alpha \in \{0.4, 0.6, 0.8\}$ and the use of dynamic scaling $\beta \in \{\text{yes}, \text{no}\}$. We experimented with distilling to different transformer layers and found that aligning with representations from the final layer yields the best results. The number of epochs allocated to the distillation phase is dataset-specific; however, in most cases it accounts for approximately 50\% of the total training epochs, after which the model is fine-tuned on the recommendation task alone. All hyperparameters were selected based on performance on validation set, and final results are averaged over 5 runs with different random seeds.

%% file: sections/4_results.tex
\section{Results}

\begin{table}[b]
    \centering
    \caption{LLM distillation performance.
    Results are averaged over 5 random seeds.}
    \resizebox{\textwidth}{!}{
    \begin{tabular}{l|cc|cc|cc|cc}
        \hline
        & \multicolumn{2}{c|}{Beauty} & \multicolumn{2}{c|}{ML-20M} & \multicolumn{2}{c|}{Kion} & \multicolumn{2}{c}{Amazon M2} \\
        & NDCG@10 & Recall@10 & NDCG@10 & Recall@10 & NDCG@10 & Recall@10 & NDCG@10 & Recall@10 \\
        \hline
        {SASRec} & 0.0106 & 0.0217 & 0.0453 & 0.0781 & 0.0585 & 0.1135 & 0.3647 & 0.5373 \\
        & $\pm$0.0004 & $\pm$0.0013 & $\pm$0.0062 & $\pm$0.0093 & $\pm$0.0009 & $\pm$0.0009 & $\pm$0.0071 & $\pm$0.0123 \\
        \hline
        \textbf{SASRec +} & \textbf{0.0111} & \textbf{0.0228} & \textbf{0.0479} & \textbf{0.0819} & \textbf{0.0597} & \textbf{0.1145} & \textbf{0.3761} & \textbf{0.5414} \\
        \textbf{LLM Distillation} & $\pm$0.0002 & $\pm$0.0004 & $\pm$0.0007 & $\pm$0.0019 & $\pm$0.0005 & $\pm$0.0007 & $\pm$0.0041 & $\pm$0.0051 \\
        \hline
        {Uplift (\%)} & +4.90\% & +5.20\% & +5.62\% & +4.74\% & +2.02\% & +0.94\% & +3.14\% & +0.75\% \\
        \hline
        \addlinespace
        \addlinespace
        \hline
        {BERT4Rec} & 0.0051 & 0.0102 & 0.0623 & 0.1088 & 0.0574 & 0.1101 & 0.2699 & 0.4230 \\
        & $\pm$0.0004 & $\pm$0.0004 & $\pm$0.0001 & $\pm$0.0008 & $\pm$0.0011 & $\pm$0.0018 & $\pm$0.0018 & $\pm$0.0013 \\
        \hline
        \textbf{BERT4Rec +} & \textbf{0.0061} & \textbf{0.0126} & \textbf{0.0628} & \textbf{0.1099} & \textbf{0.0596} & \textbf{0.1149} & \textbf{0.2727} & \textbf{0.4267} \\
        \textbf{LLM Distillation} & $\pm$0.0005 & $\pm$0.0008 & $\pm$0.0014 & $\pm$0.0029 & $\pm$0.0009 & $\pm$0.0008 & $\pm$0.0020 & $\pm$0.0019 \\
        \hline
        {Uplift (\%)} & +19.61\% & +23.53\% & +0.80\% & +1.01\% & +3.83\% & +4.36\% & +1.04\% & +0.87\% \\
        \hline
    \end{tabular}}
    \label{tab:performance}
\end{table}

\textbf{Distillation quality}.
Table \ref{tab:performance} presents comparison of sequential recommender models with and without LLM-based distillation. The results indicate that both SASRec and BERT4Rec benefit from LLM distillation. 
The effect is most pronounced when performance of the initial transformer model  is weak (as on Beauty, where BERT4Rec improves by nearly 20\%) and is less pronounced on datasets where baseline shows already high metrics (such as ML-20M). SASRec also demonstrates consistent improvements, with gains in NDCG@10 between 2.02\% and 5.62\%, indicating that even strong baseline models can benefit from LLM-derived user knowledge. Given SASRec's superior baseline performance across datasets, we selected it as the primary model for subsequent experiments.

On the Beauty dataset, the ablation in Table \ref{tab:alpha_bets} shows that without dynamic scaling ($\beta=no$), higher distillation weight ($\alpha=0.8$) performs best, whereas enabling automatic scaling shifts the optimum towards lower static $\alpha$, confirming that dynamic scaling effectively balances reconstruction and recommendation losses without manual empowering $\mathcal{L}_{distill}$ (best $\alpha=0.4$).

\begin{table}[t]
\begin{minipage}[t]{.25\textwidth}
    \centering
    \caption{\centering $\alpha,\beta$ ablation on Beauty 
    }
    \label{tab:ndcg_baseline}
    \resizebox{0.55\textwidth}{!}{%
    \begin{tabular}{|c|c|c|}
        \hline
        $\beta$ & $\alpha$ & NDCG@10 \\
        \hline
        \multirow{3}{*}{no}  & 0.4 & 0.0105\\
                             & 0.6 & 0.0108 \\
                             & 0.8 & 0.0109 \\
        \hline
        \multirow{3}{*}{yes} & 0.4 & 0.0111 \\
                             & 0.6 & 0.0106 \\
                             & 0.8 & 0.0106\\
        \hline
    \end{tabular}
    \label{tab:alpha_bets}
    }
\end{minipage}
\begin{minipage}[t]{.8\textwidth}
    \centering
    \caption{\centering Comparison of proposed approach \\and IDGenRec}
    \label{tab:ndcg_baseline}
    \resizebox{0.8\textwidth}{!}{%
    \begin{tabular}{lcccc} 
        \toprule
        Method & Beauty & ML-20M & Kion & Amazon M2  \\ 
        \midrule
        IDGenRec & \textbf{0.0114} & 0.0313 & 0.0452 & 0.1001 \\
        SASRec & 0.0106 & 0.0453 & 0.0585 & 0.3647  \\
        SASRec + LLM Distillation & 0.0111 & \textbf{0.0479} & \textbf{0.0597} & \textbf{0.3761} \\
        \bottomrule
    \end{tabular}
    }
\end{minipage}
\end{table}

\textbf{Comparison with LLM-based recommender}.
Table~\ref{tab:ndcg_baseline} highlights trade-offs between accuracy and computational cost across domains. Here, IDGenRec achieves the best performance on Beauty (NDCG@10 = 0.0114), consistent with the original report \cite{tan2024idgenrec}, while our method follows closely (0.0111) yet clearly surpasses vanilla SASRec (0.0106). Conversely, our approach outperforms IDGenRec on ML-20M, Kion, and Amazon M2, reflecting the latter’s reliance on semantic ID generation from item metadata—an advantage that diminishes when metadata are sparse or noisy. By distilling user-centric knowledge, our model maintains stable gains regardless of domain characteristics.

\begin{table*}[t]
    \centering
    \caption{Comparison of Training (for one epoch) and Inference Time (for test set) with IDGenRec, seconds}
    \resizebox{\textwidth}{!}{%
    \begin{tabular}{l|cc|cc|cc|cc}
        \hline
        \multirow{2}{*}{Method} & \multicolumn{2}{c|}{Beauty} & \multicolumn{2}{c}{ML20M} & \multicolumn{2}{c}{Kion} & \multicolumn{2}{c}{Amazon M2} \\
        \cline{2-9}
        & Training $\downarrow$ & Inference $\downarrow$ & Training $\downarrow$ & Inference $\downarrow$ & Training $\downarrow$ & Inference $\downarrow$ & Training $\downarrow$ & Inference $\downarrow$ \\
        \hline
        IDGenRec & 37.11 & 120.38 & 132.12 & 840.01 & 80.04 & 342.03 & 1529.45 & 7492.11 \\
        SASRec & 25.56 & 2.37 & 68.30 & 4.37 & 30.93 & 1.12 & 663.26 & 41.07\\
        SASRec + LLM Distillation & 26.92 & 2.37 & 71.34 & 4.37 & 31.81 & 1.12 & 833.36 & 41.07\\
        \hline
    \end{tabular}}
    \label{tab:time}
\end{table*}

Table \ref{tab:time} further underscores our method’s efficiency advantages. Training overhead increases only by 5–25\% over SASRec, compared to IDGenRec’s 1.5–2.3× longer training time. At inference, our approach matches SASRec’s latency, while IDGenRec slows by 50–180× due to beam search–based text generation. These results demonstrate that our distillation strategy combines LLM-level semantic richness with the practical efficiency of standard forward inference, making it suitable for large-scale, real-time recommender systems.

\begin{figure}[t]
  \centering
  \begin{subfigure}[t]{0.5\columnwidth}
    \centering
    \includegraphics[width=\linewidth]{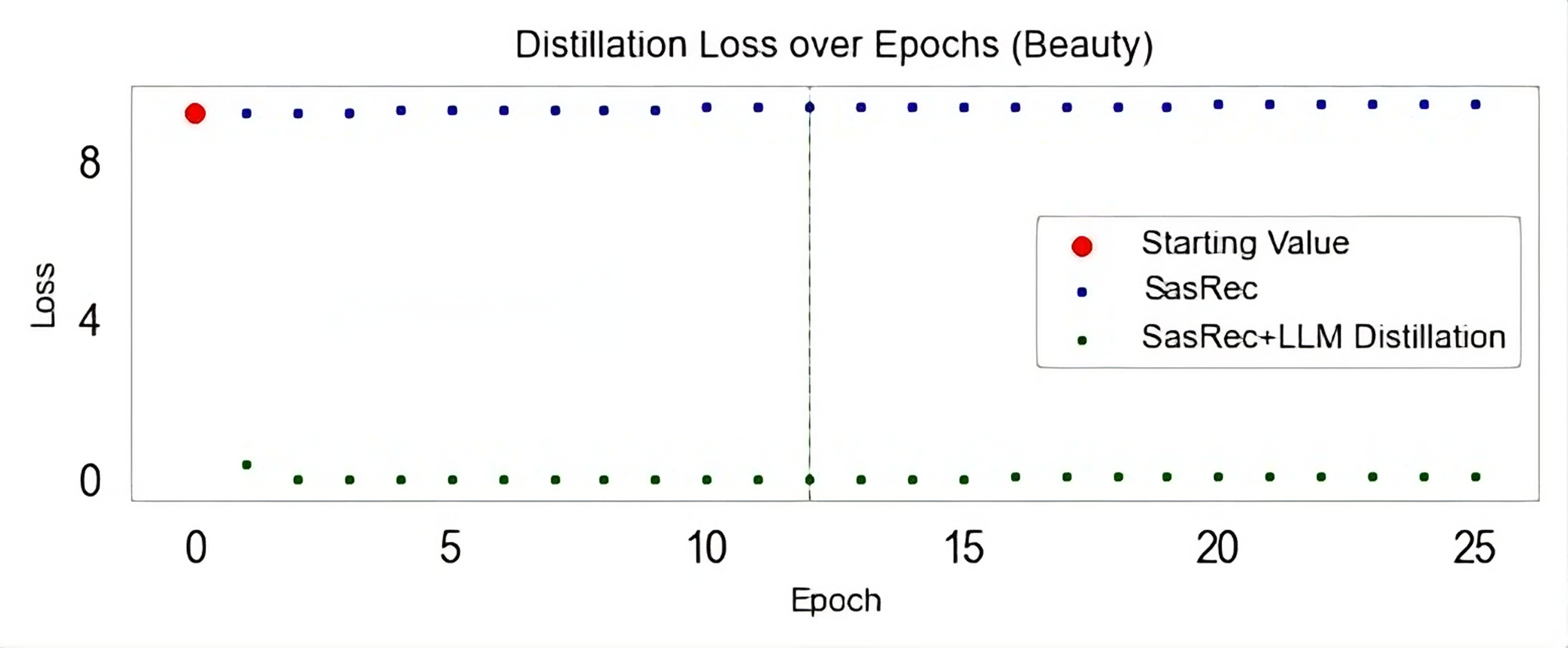}
  \end{subfigure}%
  \hfill
  \begin{subfigure}[t]{0.5\columnwidth}
    \centering
    \includegraphics[width=\linewidth]{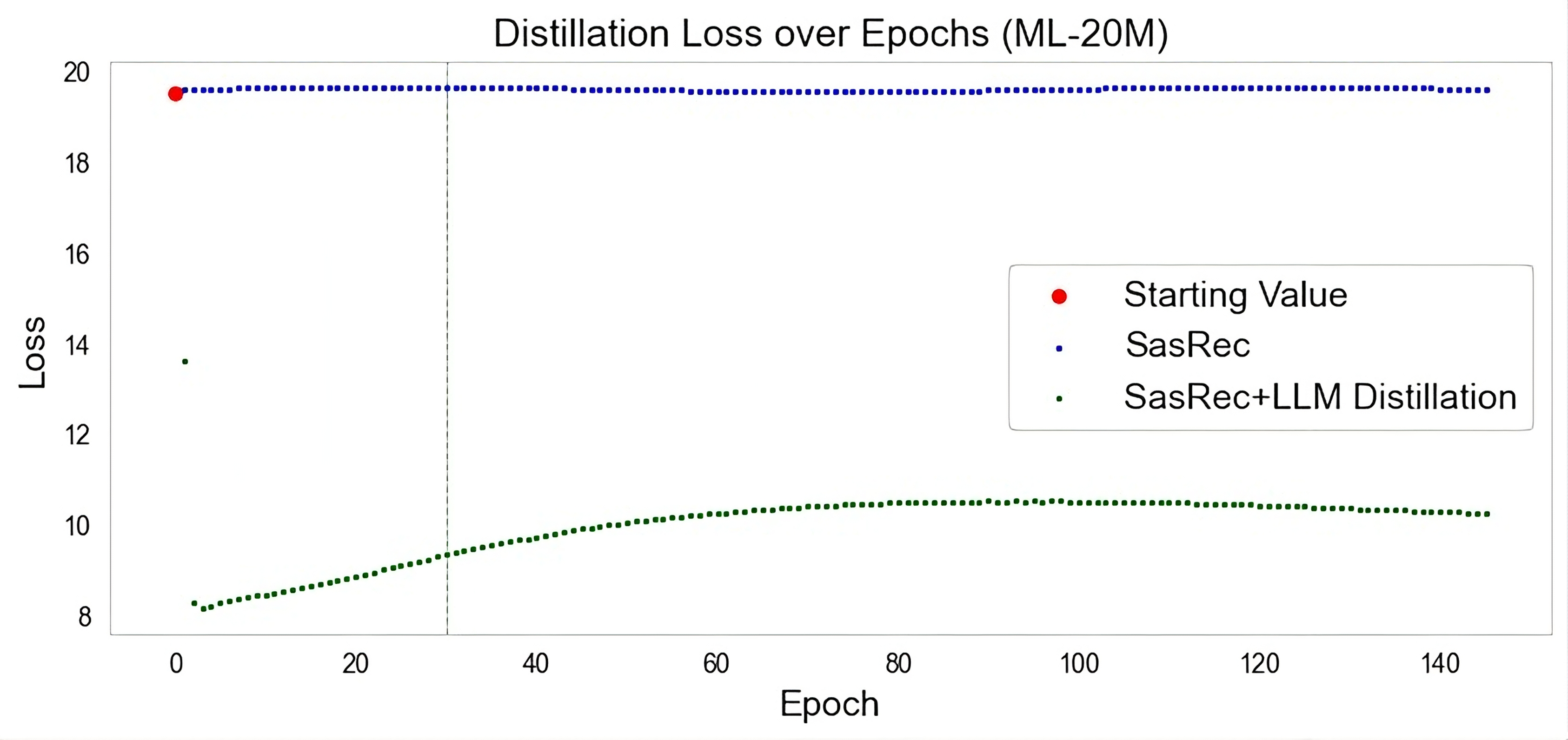}
  \end{subfigure}
  \caption{Distillation loss $\mathcal{L}_{distill}$ trajectories across training epochs. 
  The green vertical line marks the transition between training phases. 
  }
  \label{fig:loss_comparison}
\end{figure}

\textbf{Auxilary loss behaviour}.
Figure~\ref{fig:loss_comparison} compares reconstruction loss trajectories for vanilla SASRec and our two-phase LLM-distilled variant on Beauty and ML-20M. The LLM-distilled model quickly converges to low reconstruction loss and remains stable after the phase transition, indicating successful integration of LLM-derived user knowledge. While the vanilla model shows persistently high loss, the distilled model preserves reconstruction ability even after the distillation signal is removed, demonstrating that the learned representations capture generalizable structure in user behavior.

%% file: sections/5_conclusion.tex
\section{Conclusion}

This paper introduced a novel knowledge distillation framework that transfers user preference insights from pre-trained large language models to sequential recommendation systems without incurring inference-time computational overhead. The method effectively leverages LLM semantic understanding while remaining efficient for large-scale deployment. The source code of our experiments is publicly available\footnote{\url{https://github.com/sb-ai-lab/ECIR26_Pre-trained_LLMs_Meet-Sequential_Recommenders}}.

Experiments on four diverse datasets show consistent gains over strong baselines, up to 23.5\% in Recall@10, and demonstrate stable retention of LLM-derived knowledge during training. Our approach matches the performance of existing LLM-based models while achieving significantly faster inference (up to 60×) and training (2.5×) speeds.

This work highlights a practical path toward integrating LLM capabilities into recommender systems and other domains where semantic richness and efficiency must coexist. Future work includes exploring broader architectures and more advanced user profile generation, such as incorporating negative feedback.